# Experimental demonstration of three-dimensional broadband underwater acoustic carpet cloak

Yafeng Bi[1,2], Han Jia[1,2,3,a)], Zhaoyong Sun[1,2], Yuzhen Yang[1,2], Han Zhao[1,2] and Jun Yang[1,2,3,a)]

[1] Key Laboratory of Noise and Vibration Research, Institute of Acoustics, Chinese Academy of Sciences, Beijing 100190, People's Republic of China

[2] University of Chinese Academy of Sciences, Beijing 100049, People's Republic of China

[3] State Key Laboratory of Acoustics, Institute of Acoustics, Chinese Academy of Sciences, Beijing 100190, People's Republic of China

**Abstract**

We present the design, architecture and detailed performance of a three-dimensional (3D) underwater acoustic carpet cloak (UACC). The proposed system of the 3D UACC is an octahedral pyramid which is composed of periodical steel strips. This underwater acoustic device, placed over the target to hide, is able to manipulate the scattered wavefront to mimic a reflecting plane. The effectiveness of the prototype is experimentally demonstrated in an anechoic tank. The measured acoustic pressure distributions show that the 3D UACC can work in all directions in a wide frequency range. This experimental verification of 3D device paves the way for guidelines on future practical applications.

---

a) Author to whom correspondence should be addressed. Electronic addresses: hjia@mail.ioa.ac.cn; jyang@mail.ioa.ac.cn

Invisibility acoustic cloaks are devices which can control the propagation of acoustic waves and hide the signature of the objects.[1-5] Among all these cloak models, carpet cloak has received considerable attention so far.[6-15] This device can restore the wavefront as if the wave is reflected from a plane, so that the cloaked target would be indistinguishable from the reflecting surface.

It is well understood that the parameters of the carpet cloak can be calculated by the transformation acoustics. By utilizing proper mapping rules, the required mass density and bulk modulus of the device can be obtained. The quasiconformal mapping is firstly proposed in designing the carpet cloak. However, its inhomogeneous parameters and large volume will lead to complex design and fabrication process.[6-10] As a result, researchers turn to a more feasible approach — the carpet cloak with linear transformation.[11-15] The linear mapping rule brings homogeneous parameters with reasonable anisotropy, which can be realized by acoustic metamaterial.[16, 17]

By using the layered perforated plates, Cummer *et al.* designed a 2D carpet cloak in the air host and confirmed its effectiveness by measuring the scattered acoustic pressure fields.[13] Then, this concept was extended to 3D. A square pyramid carpet cloak in the air host was fabricated and demonstrated experimentally.[15] However, because of the difficulty in obtaining the required parameters, the underwater acoustic carpet cloak (UACC) remained on the stage of simulation for some time.[18, 19]

Recently, we have designed a 2D UACC.[20] By introducing a scaling factor, the structure of the UACC, which is just composed of layered brass plates, is greatly simplified with some impedance mismatch. Measured acoustic field distributions verified the cloak effect in a wide frequency range. In this article, we further extend the UACC to 3D. The scaling factor is also considered in our design and the UACC is simply made up of periodical steel strips. The measured acoustic pressure distributions demonstrate that the designed 3D device can work in all directions in a wide frequency range.

The proposed system of the 3D UACC is depicted in Fig. 1(a). It is an octahedral pyramid which contains two spaces. One is a small octahedral pyramid (the 3D space in *aOb*) placed on a reflecting surface. It is the cloaked space which can be used to hide objects. The other is the UACC (the 3D space in *abc*) which covers the cloaked space. The whole system height *c* is 35 cm, half the octahedral pyramid base length *b* is 28.3 cm and the height of the cloaked space *a* is 7.1 cm. The linear transformation rule used in the design maps the cloaked space into a flat plane and makes the objects hidden in the cloaked space undetectable. By using the transformation acoustics theory,[1, 2, 13] required density and bulk modulus of the 3D UACC have been calculated as follows: $\rho_{11}^{pr} = 3.38\rho_0$, $\rho_{22}^{pr} = 1.75\rho_0$, $\rho_{33}^{pr} = 1.61\rho_0$, $\kappa^{pr} = 1.75\kappa_0$, where $\rho_0$ and $\kappa_0$ are density and bulk modulus of the background fluid and superscript *pr* represents the principal direction. Here, to make the UACC more realizable, a scaling factor $\omega = 0.43$ has been introduced in the parameter calculation.[20] This scaling factor moderately modulates the impedance of the device without changing the distribution of sound velocity. So it has limited impact on the camouflage effect.

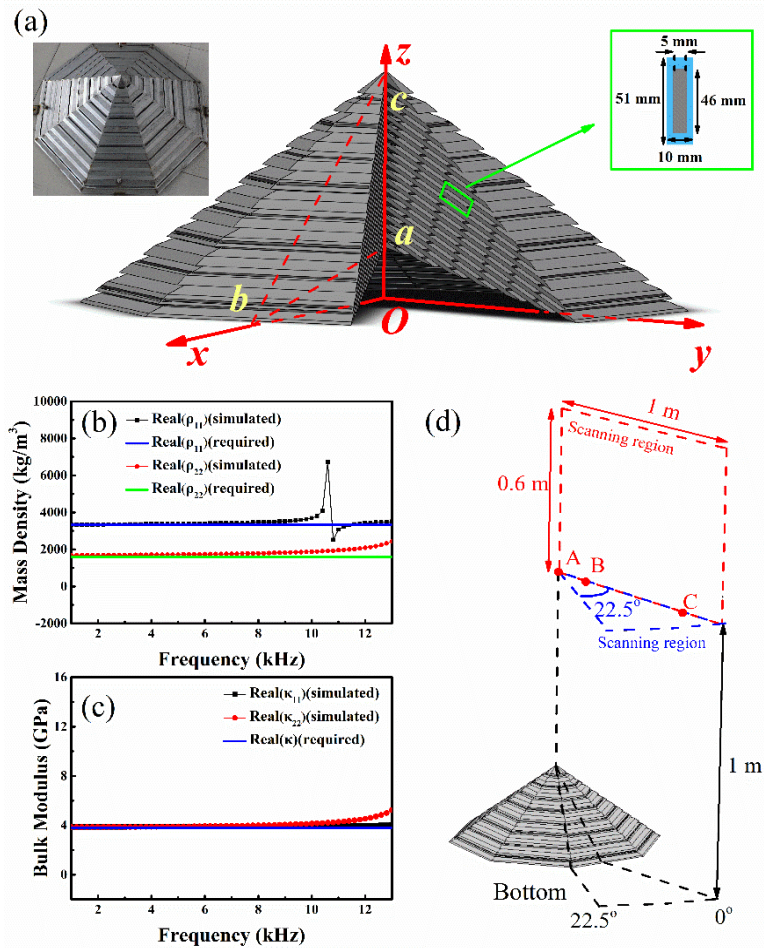

FIG. 1. (a) Schematic view of the 3D UACC. A typical unit cell is enlarged and presented in the green frame and the photograph of the 3D UACC sample is shown in the left inset in (a). (b) The simulated and required effective mass density of the unit cell. (c) The simulated and required effective bulk modulus of the unit cell. (d) Schematics of the measured areas. Two scanned areas are shown as the red rectangle frame and blue sector frame, respectively. A, B and C are three particular points to extract the time domain signals.

Then, we use the artificial structures to realize the required acoustic parameters and fabricate the prototype of 3D UACC. Since the system can be split into eight identical parts, we just need to take a triangular pyramid into consideration in the design process. The steel strip array is chosen to establish the structure. From Fig. 1(a), it can be observed that the cross section of the triangular pyramid is composed of periodical steel strips with a rectangle lattice. A unit cell is enlarged and exhibited in green frame. It can be seen that the size of the lattice is 51 mm by 10 mm, while the width and thickness of these strips are 46 mm and 5 mm respectively. By varying the length with the space positions, these steel strips form a triangular pyramid. We can splice eight identical triangular pyramids and integrate them into a whole prototype. The steel strips (density: 7850 kg/m$^3$; Young's modulus: 205 GPa; Poisson's ratio: 0.28) are separated by water (density: 1000 kg/m$^3$; acoustic velocity: 1480 m/s) in the proposed system, so that the influence of shear wave between these solid strips can be minimized. In most cases, this unit cell can be approximately regarded as fluid in long wavelength regime.

By using the well-developed retrieval method in simulation,[21] the effective acoustic parameters of the unit cell can be calculated. The results are presented in Fig. 1(b) and (c). Figure 1(b) shows the simulated effective mass density versus frequency. The simulated mass densities in two principal axes are plotted as symbols, while the required values of the UACC are plotted as lines for comparison. The different values in two directions indicate that the mass density tensor is anisotropic. The fluctuation of $\rho_{11}$ around 11 kHz is caused by the excitation of a bending vibration mode of the steel strips. However, the excitation of this vibration mode is direction-dependent and just in a narrow band, so it has little influence on the effectiveness of the 3D UACC. The simulated (symbols) and required (line) bulk moduli in two principal axes are presented in Fig. 1(c). They keep stable in the entire band of interest, and both bulk moduli in two directions are nearly the same. As the result, the proposed unit cell has effective isotropic bulk modulus and anisotropic mass densities, which match well with the parameters required by the UACC.

To demonstrate the effectiveness of the designed 3D UACC, a series of experiments are conducted in an anechoic tank. The sample is submerged under 5 meters of water. An omnidirectional cylindrical transducer is placed above the sample (about 3 m away from the bottom of the sample). In order to distinguish the scattered wave from the incident wave, a series of short Gaussian pulses (7 kHz to 13 kHz with 1 kHz step) are used as the detection signals in experiment. The propagation of the wave is measured in time-domain using two hydrophones (Type 8103, B&K): one is fixed near the transducer as the monitoring hydrophone while the other scans the measuring region step by step. All the emitting and receiving acoustic signals are analyzed by a multianalyzer system (Type 3560, B&K).

Firstly, a rectangle area marked with red dash lines in Fig. 1(c) is measured to display the distribution of scattered wave. It is 1 m wide and 0.6 m high. The bottom is 1 m away from the reflecting plane. In measuring the red area, the scanning hydrophone moves on a square grid of 2 cm to ensure at least five measurement points per wavelength. To better exhibit the cloaking effect, the pressure fields were measured in three situations: the soft plane (air sealed by thin steel plates) placed underwater, the soft target (air octahedral pyramid sealed by thin steel plates) placed underwater and the cloaked soft target placed underwater, respectively.

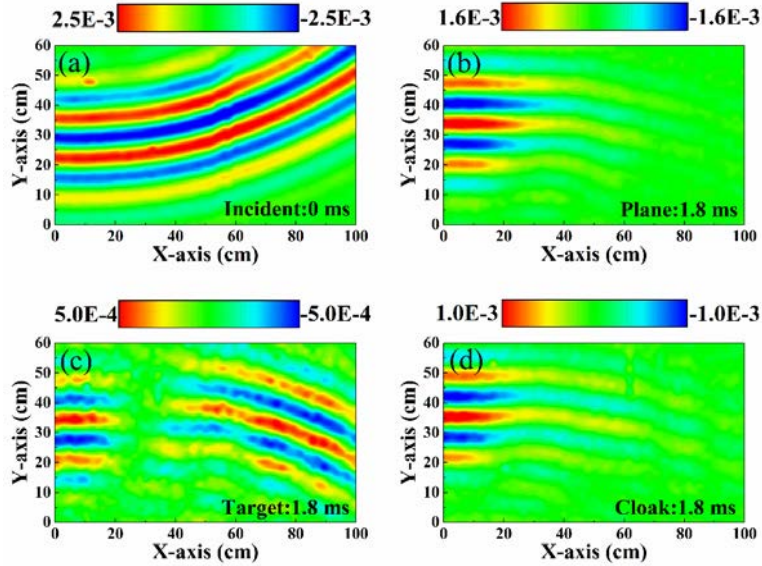

FIG. 2. Measured acoustic pressure fields in the rectangle frame at 10 kHz. (a) The incident pressure field at 0 ms. (b) – (d) The scattered pressure fields at 1.8 ms from the soft plane (b), the soft target (c) and the cloaked soft target (d), respectively.

The measured incident and scattered acoustic fields at 10 kHz are shown in Fig. 2. In Fig. 2(a), the transducer produces a Gaussian pulse whose center frequency is 10 kHz. The omnidirectional wave comes from the upper left corner, then spreads around and propagates into the scanning area. After 1.8 ms, the wave is scattered by the objects and propagates from the bottom to the top. Figure 2(b) shows the reflection of the soft plane. The wave packet keeps a Gaussian shape and focuses on the backscattering direction. In Fig. 2(c), the wave is scattered by the soft target. Due to the slopes of the target, acoustic wave impinges on the surfaces obliquely and reflects to the opposite side. Therefore, most of the scattered wave inclines to sides of the target. For the finite length of the slope, the sharp edge of the soft target will cause extra scattering, which generates the reflected wave propagating along the backscattering direction. However, the amplitude of the backscattering wave from the target is much smaller than that from the plane. The scattered acoustic field of the target is obviously different with that of the soft plane. In contrast, after covering the target with 3D UACC, the scattered wave returns to the backscattering direction again (Fig. 2(d)). Both the phase and the amplitude of the scattering wave are much closer to those from the soft plane. The scattering field of the cloaked object is significantly similar with the scattering field of the soft plane. All these phenomena indicate that the acoustic signature of the target is almost cancelled and cloaked target mimics the soft plane successfully.

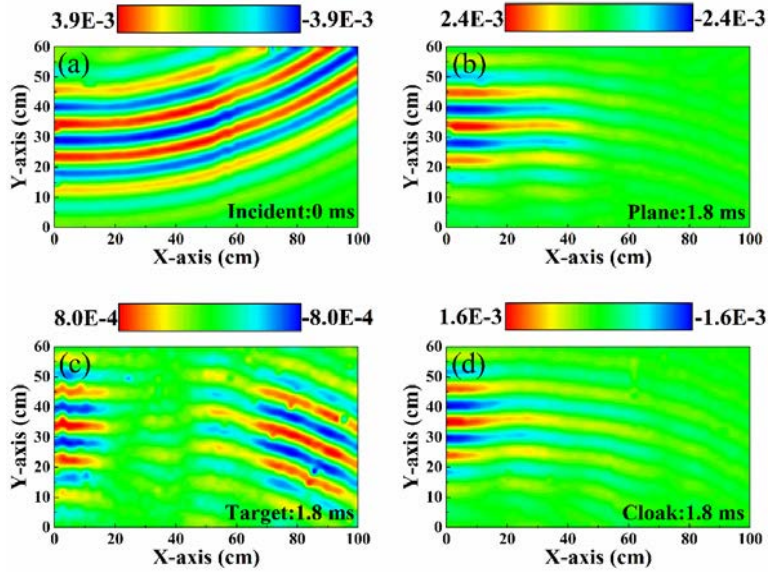

FIG. 3. Measured acoustic pressure fields in the rectangle frame at 13 kHz. (a) The incident pressure field at 0 ms. (b) – (d) The scattered pressure fields at 1.8 ms from the soft plane (b), the soft target (c) and the cloaked soft target (d), respectively.

Similar acoustic fields at 13 kHz are also provided in Fig. 3. Comparing the phase, amplitude and propagation direction in these three acoustic pressure fields, it is obviously that the distributions of the scattered wave are recovered after covering the 3D UACC.

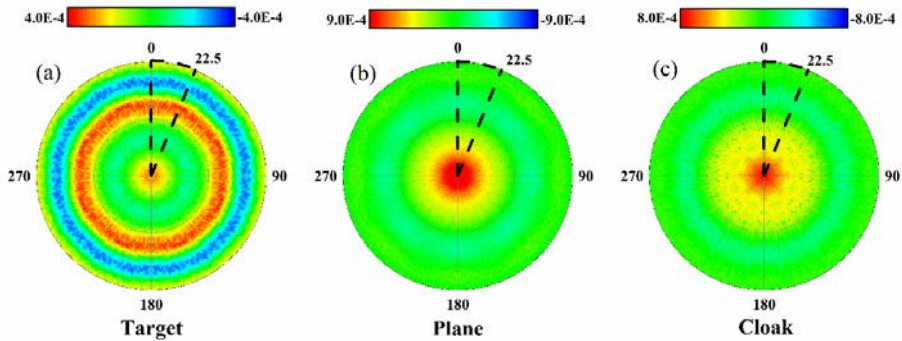

FIG. 4. Measured scattered acoustic pressure fields in the sector frame at 8 kHz from the soft plane (b), the soft target (c) and the cloaked soft target (d), respectively.

To demonstrate the overall camouflage effect of the 3D UACC, we further measure a sector area marked with blue dash lines, as shown in Fig. 1(d). This measured area can show the effectiveness of the 3D UACC at a view of overlooking sight. In measuring the blue sector area, the scanning hydrophone moves on a length of 2 cm along the radius direction and 1 degree along the angle direction. For the symmetry of the sample, we have just measured a sector of 22.5 degrees. In Fig. 4, the measured sector fields have been replicated and extended to a circular area for better demonstrations. These three panels display the acoustic pressure fields at 1.6 ms when the scattered wave comes into the scanned region. In Fig. 4(a), the annular distribution of the scattered wave indicates that the wave is diffusing out from the target. While in Fig. 4(b), the scattered wave from the soft plane is focusing on the center of the circle, which indicates that the wave is propagating along the backscattering direction. After covering the 3D UACC (Fig. 4(c)), the scattered wave is also focusing on the center of the circle. Besides, both the scattered wave from the soft plane and cloaked target are positive pressure (red color) at the same time, which also demonstrates that their phase matches well with each other. In Fig. 4(c), it can also be observed that the acoustic pressure is evenly distributed along the angle direction, which demonstrates that the ridge of the 3D UACC has little impact on the camouflage effect.

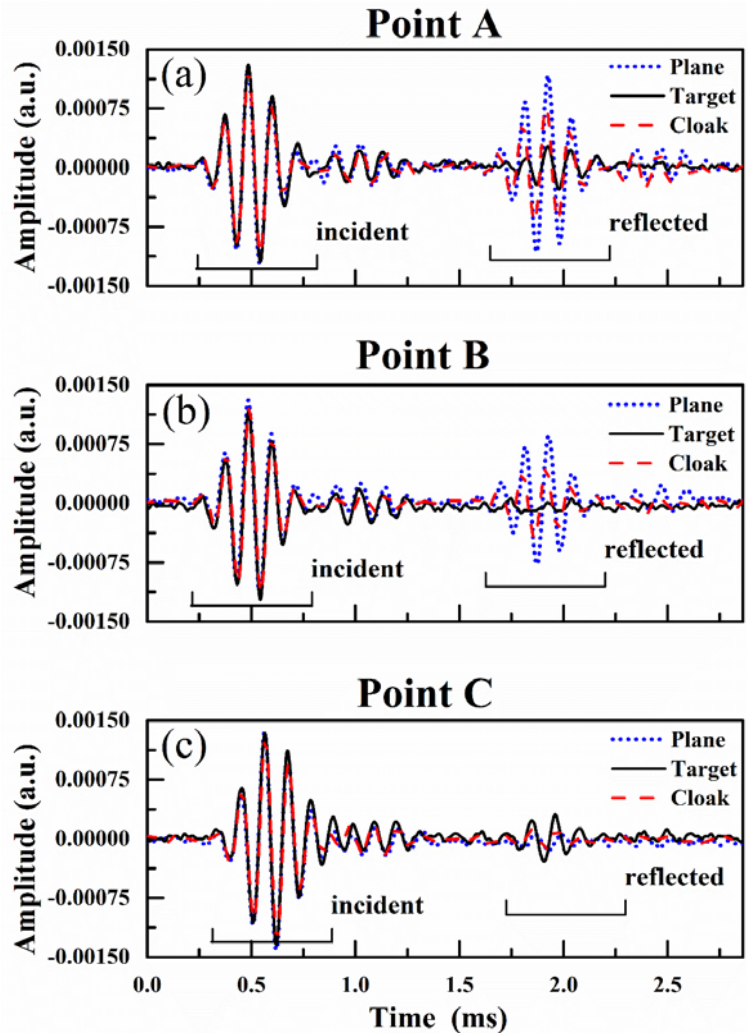

FIG. 5. Extracted time domain signals at 8 kHz. The locations of A, B and C are specified in Fig. 1(d). Each plot shows the incident wave and the scattered wave for three scenarios: the soft plane (shown as blue dotted lines), the soft target (shown as black solid lines), and the cloaked soft target (shown as red dash lines).

Times domain signals at three special positions are also extracted to exhibit the performance of the 3D UACC. Three points A(0,100), B(20,100) and C(80,100) (shown in Fig. 1(d)) are corresponding to the point above the sample, the point in the backscattering region and the point in the direction where the scattered wave from the target spread, respectively. The signals at 8 kHz are presented in Fig. 5. Each panel contains signals obtained with the soft plane (blue dotted lines), bare soft target (black lines) and cloaked target (red dash lines), respectively. It can be seen that the incident waves in each panel are nearly identical. In Fig. 5(a) and (b), for these two points are in the backscattering direction, the amplitude of the reflected wave from the soft plane is large while the amplitude of the scattered wave from the soft target is small. However, after covering the 3D UACC on the target, the amplitude increases significantly and the phase also becomes matched with that of the soft plane. In Fig. 5(c), the point deviates from the backscattering direction, so the reflected wave from the soft plane is much weaker than that from the target. The same results are also observed after covering the 3D UACC on the target.

Moreover, the reduced total radar cross section (RCS) is used to quantitatively evaluate the camouflage effect of the 3D UACC.[22] By calculating the difference between two acoustic pressure fields, it can be obtained:

$$\sigma_{reduced} = \sigma_{cloaked}/\sigma_{uncloaked} = \oint_\Omega \left( |P_{cloaked,scat}|^2 / |P_{uncloaked,scat}|^2 \right) d\Omega$$
$$P_{cloaked,scat} = P_{cloaked,tot} - P_{plane,tot}$$
$$P_{uncloaked,scat} = P_{uncloaked,tot} - P_{plane,tot} \tag{1}$$

Where $P_{cloaked,tot}$, $P_{uncloaked,tot}$ and $P_{plane,tot}$ are the measured acoustic pressure fields at 1.8 ms when the scattered wave propagates into the scanning region. The waves are scattered from the cloaked soft bump ($P_{cloaked,tot}$), the bare soft target ($P_{uncloaked,tot}$) and the soft plane ($P_{plane,tot}$), respectively. Obviously, for the cloaking effect, the smaller value is better.

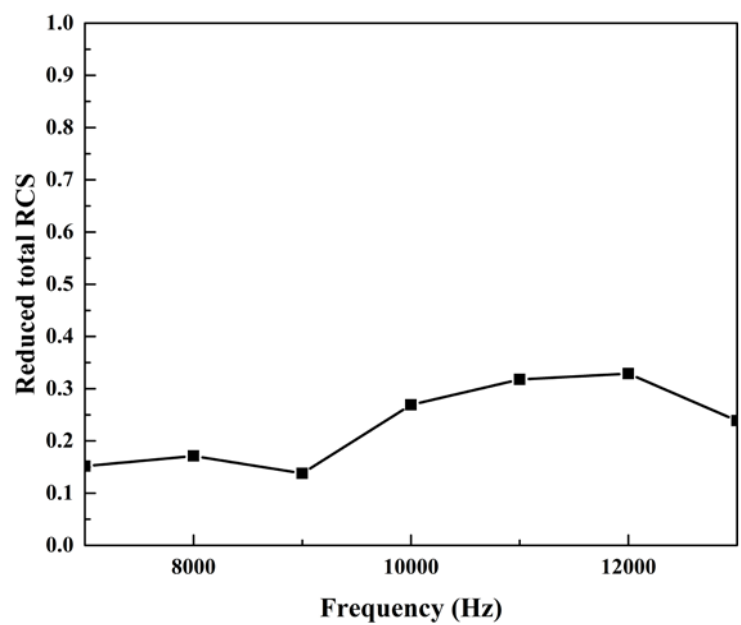

FIG. 6. Reduced total RCS from 7 kHz to 13 kHz.

The reduced total RCS from 7 kHz to 13 kHz is shown in Fig. 6. All the values are lower than 0.33, which further confirm the validity of the designed 3D UACC in a wide frequency range.

In conclusion, we realize a 3D UACC which can mimic a reflecting plane and hide the information of the 3D target. The required parameters, which contain anisotropic mass densities and isotropic bulk modulus, are obtained through steel strips array surrounded by water. The subwavelength steel-water unit cell can be regarded as fluid in long wavelength regime. Meanwhile, it ensures the carpet cloak as a broadband device. Then the performance of the carpet cloak is assessed experimentally by measuring the acoustic pressure fields in an anechoic tank. The results confirm the expected behavior and demonstrate the effectiveness and omnidirectionality of the designed 3D UACC. We believe that the verification of 3D UACC gives more direct guidance on the future practical applications of acoustic metamaterials.

## Acknowledgments

This work was supported by the Youth Innovation Promotion Association of CAS (Grant No. 2017029), the IACAS Young Elite Researcher Project (Grant No. QNYC201719) and the National Natural Science Foundation of China (Grant No. 11304351, 1177021304).